\documentclass[11pt]{article}
\usepackage{times}
\usepackage{graphics}
\usepackage{color}
\usepackage{amssymb}
\usepackage{amsmath}
\usepackage{ifthen}
\usepackage{calc}
\usepackage{float}

\newtheorem{theorem}{Theorem}

\newcommand{\QQ}[1]{}

\newlength{\Ainlength}
\newlength{\Ainindent}
\newlength{\Aintemp}
\newcommand{\Ain}[1]{\setlength{\Aintemp}{\Ainindent}\addtolength{\Aintemp}{#1\Ainlength} \hspace*{\Aintemp}}

\renewcommand{\vec}[1]{\ensuremath{\mathbf{#1}}}
\newcommand{\data}{\ensuremath{(\vec{y},\vec{w})}}
\newcommand{\err}[2]{\ensuremath{\mathsf{err}^{#1}\ifthenelse{\equal{#2}{}}{}{(#2)}}}

\newcommand{\Err}[1]{\ensuremath{\mathsf{e}\ifthenelse{\equal{#1}{}}{}{(#1)}}}
\newcommand{\Errj}[1]{\ensuremath{\mathsf{e^\prime}\ifthenelse{\equal{#1}{}}{}{(#1)}}}
\newcommand{\jopt}[1]{\ensuremath{\mathsf{j_{min}}\ifthenelse{\equal{#1}{}}{}{(#1)}}}
\newcommand{\inter}[2]{\ensuremath{[#1\!:\!#2]}}
\newcommand{\epslow}{\ensuremath{\epsilon_\mathrm{low}}}
\newcommand{\epshigh}{\ensuremath{\epsilon_\mathrm{high}}}
\newcommand{\epsopt}{\ensuremath{\epsilon_\mathrm{opt}}}
\newcommand{\epsinterval}[1]{\ensuremath{(\epslow,\epshigh#1}}
\newcommand{\bibshrink}{\vspace{-0.04in}}

\floatstyle{plain}
\newfloat{algorithm}{tbp}{algorithm}

\floatname{algorithm}{Algorithm}

\begin{document}

\begin{center}
\textbf{\Large An Algorithm for $\mathbf{L_\infty}$ Approximation by Step Functions}
\medskip

{\large Quentin F. Stout}
\smallskip

Computer Science and Engineering\\
University of Michigan\\
Ann Arbor, MI 48109-2121\\
qstout@umich.edu~~~~~ +1 734.763.1518
\end{center}
\vspace{-0.17in}

\subsubsection*{Abstract}
An algorithm is given for determining an optimal $b$-step approximation of weighted data, where the error is measured with respect to the $L_\infty$ norm.
For data presorted by the independent variable the algorithm takes $\Theta(n + \log n \cdot b(1+\log n/b))$ time and $\Theta(n)$ space.
This is $\Theta(n \log n)$ in the worst case and $\Theta(n)$ when $b = O(n/\log n \log\log n)$.
A minor change determines an optimal reduced isotonic regression in the same time and space bounds,
and the algorithm also solves the $k$-center problem for 1-dimensional weighted data.

\noindent
\textbf{Keywords}: step function approximation; reduced isotonic regression; variable width histogram; weighted k-center; interval tree of bounded envelopes

\vspace*{-0.055in}
\section{Introduction}
\vspace*{-0.055in}

Step functions are a fundamental form of approximation, arising in variable width histograms,
databases, segmentation,
approximating sets of planar points, piecewise constant approximations, etc.
Here we are interested in $L_\infty$ stepwise approximation of weighted data.
By weighted data \data\ on $1\ldots n$ we mean values $(y_i,w_i)$,
$1 \leq i \leq n$, where $y_i$ is an arbitrary real number and $w_i$ (the weight) is a positive real number.
For integers $i \leq j$ let \inter{i}{j} denote $i \ldots j$.
A function $f$ on \inter{1}{n} is a
\textit{$b$-step function} iff there are indices $j_1 = 1 < j_2 < \ldots < j_{b+1} =n+1$ and real values $C_k$, $k \in \inter{1}{b}$, such that $f_i = C_k$ for $i \in \inter{j_k}{j_{k+1}\!-\!1}$.
$f$ is an \textit{optimal $L_\infty$ $b$-step approximation of \data} iff it minimizes the weighted $L_\infty$ error, $\max\{w_i\cdot|f_i-y_i| : i\in\inter{1}{n}\}$, among all $b$-step functions.
Since a step can be split into smaller ones, we do not differentiate between ``$b$ steps'' and ``no more than $b$ steps''.

Many algorithms have been developed for $L_\infty$ $b$-step regression~\cite{ChenWangPiecewise2013,ChenWangnlogn2013,DiazBanezLinearDecide,FournierVigneronLinftyStep,FournierVigneronLinftyParametric,FulopPrillLinftyStep,GuhaShimLinftyHistogram,JQReducedIso_IF2012,KarrasetalHistogramDuality,LiuRandomizedLinftyReduced,MaysterLopezStep}.
The first $\Theta(n \log n)$ time deterministic algorithms~\cite{FournierVigneronLinftyParametric,JQReducedIso_IF2012} were decidedly impractical, relying on parametric search.
A more feasible $\Theta(n \log n)$ algorithm appeared in~\cite{ChenWangnlogn2013}.
However, the time of these algorithms does not improve when $b$ is small, which is the typical case of interest.
We present a faster algorithm that is $\Theta(n + \log n \cdot b(1+\log n/b))$ when the data is presorted by independent coordinate.

\vspace*{-0.055in}
\section{$L_\infty$ $b$-Step Approximation} \label{sec:Linfty}
\vspace*{-0.055in}

At a high-level overview, our algorithm shares aspects of those in~\cite{ChenWangPiecewise2013,FournierVigneronLinftyStep,GuhaShimLinftyHistogram}, with important differences:
\vspace*{-0.05in}
\begin{enumerate}
\item Build an interval tree to determine the regression error of an arbitrary interval if it is a single step.
\vspace*{-0.05in}
\item Use ``search in a sorted matrix'' to find the minimal possible error for a $b$-step approximation.
\end{enumerate}
\vspace*{-0.05in}
The search uses a feasibility test which is given $\epsilon$ and decides if there is a $b$-step approximation with error $\leq \epsilon$.
If there is such an approximation then the test produces one.
We incorporate important improvements to this approach: feasibility tests are used during tree construction, not just during the search; tests do not determine the minimal regression error of an interval, merely that it is sufficiently small or too large; previous searches, except the randomized version in~\cite{LiuRandomizedLinftyReduced}, were not search in a sorted matrix; and we exploit the fact that calculations at one stage of the search are related to those of the previous stage.
We will show:
\begin{theorem} \label{thm:mainresult}
Given weighted data \data, sorted by the independent coordinate, and number of steps $b$, one can determine an optimal $L_\infty$ $b$-step approximation 
in $\Theta(n + \log n \cdot b (1+\log n/b))$ time and $\Theta(n)$ space.
\end{theorem}

\vspace*{-0.05in}
Given a set \data\ of weighted values and $k \in \inter{1}{n}$, the \textit{1-dimensional weighted $k$-center problem} is to find a set $S = \{s_1, \ldots, s_k\}$ of real numbers that minimizes $\max\{d(y_i,S): i\in \inter{1}{n}\}$,
where $d(\cdot,S)$ is the weighted distance to $S$, i.e., $d(y_i,S) = \min\{w_i\cdot|y_i - s_j|: j\in \inter{1}{k}\}$.
Note that a set $S$ is an optimal $k$-center iff it is the step values of an optimal $L_\infty$ $k$-step approximation of the values in sorted order.

From now on we generally omit mention of ``$L_\infty$'' and ``optimal'' since they are implied.

\subsection{Interval Tree of Bounded Envelopes} \label{sec:LinftyEnvelopes}

For a weighted value $(y,w)$, in the $y$-$z$ plane the error of using $z \geq y$ as its regression value is given by the ray in the upper half-plane that starts at $(y,0)$ with slope $w$.
Given a set of weighted data \data, its \textit{upward error envelope} is the topmost sequence of line segments corresponding to all such rays.
For each $z$, it gives the maximum error of using $z$ as the regression value for all points $(y_i,w_i)$ where $z \geq y_i$.
The \textit{downward error envelope} uses rays in the upper half-plane starting at $(y_i,0)$ with slope $-w_i$, representing the error of using a regression value $\leq y_i$.
The intersection of the downward and upward error envelopes gives the regression value minimizing the error over the entire set, i.e., the weighted $L_\infty$ mean, and its error.

\begin{figure}

\begin{center}
\resizebox{4.4in}{!}{\includegraphics{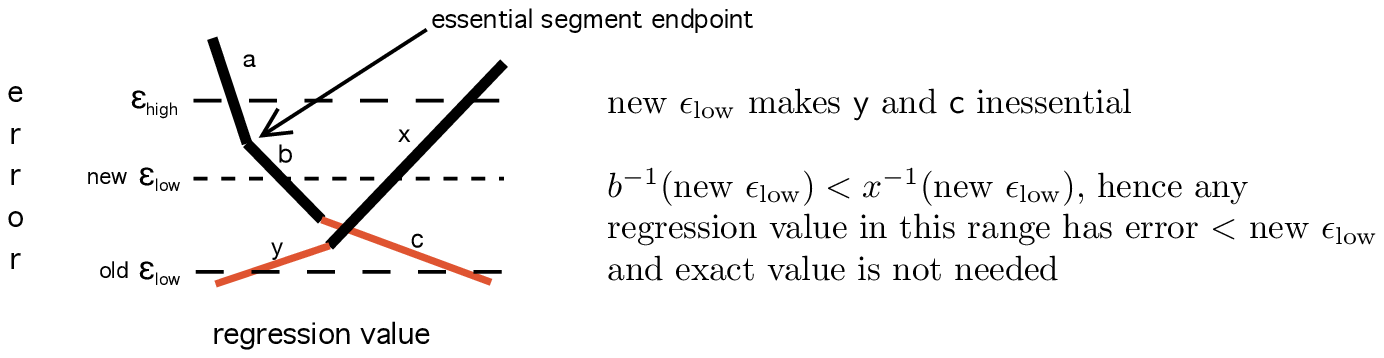}}
\end{center}
\vspace{0.08in}

\hrulefill
\vspace*{-0.55in}
\caption{Downward and upward bounded envelopes}  \label{fig:envelopes}
\vspace*{-0.0in}
\end{figure}

To simplify exposition we assume that $n$ is an integral power of 2.
A binary \textit{interval tree} has a root corresponding to the interval \inter{1}{n},
its two children correspond to \inter{1}{n/2} and \inter{n/2+\!1}{n}, 
their children represent intervals of length $n/4$, etc. 
The leaves are \inter{1}{1}, \inter{2}{2}, \ldots \inter{n}{n}.
The intervals corresponding to nodes will be called \textit{binary intervals}.

Some authors~\cite{ChenWangPiecewise2013,FournierVigneronLinftyStep,GuhaShimLinftyHistogram} used interval trees where each node 
contains the upward and downward error envelopes of the data in its interval, but most of the envelopes' segments are unnecessary.
Let $\epsopt$ be the (unknown) error of an optimal $b$-step approximation,
and suppose bounds $\epslow < \epsopt \leq \epshigh$ are known.
Then $\epsopt$ can be determined using only the segments representing errors in \epsinterval{)}.
These will be called \textit{essential segments}, and they form \textit{bounded envelopes}.
All others, the \textit{inessential segments}, are discarded.
Figure~\ref{fig:envelopes} shows how essential segments can become inessential when a better error bound is determined.
Initially $\epslow=0$ and $\epshigh=\infty$, and the algorithm continually improves these bounds.
In our interval tree each node contains its upward and downward bounded envelopes.
Throughout, the essential segments are precisely the segments in the standard, unbounded, envelopes that represent errors in \epsinterval{)}.
Thus correctness depends on properties of the standard envelopes, though timing does not.

Bounded envelopes are stored as a doubly-linked list ordered by slope.
Whenever a node is visited, by starting at both ends, inessential segments (i.e., those with no errors in \epsinterval{)}) are discarded.
The time is charged to the segments removed, not the search visiting the node.
Only $\Theta(n)$ segments are ever created, hence the total time to remove inessential ones is $\Theta(n)$.
Whenever the number of remaining segments is counted the count is only of the essential segments given the current values of \epslow\ and \epshigh.

\subsection{Feasibility Tests} \label{sec:feasibility}

Given an arbitrary interval let $U(S)$ denote the upward bounded envelope of the data in $S$, and $D(S)$ the downward bounded envelope. 
For $\epsilon \in \epsinterval{)}$ the error of making $S$ a single step is $<$, $=$, $>$ $\epsilon$ iff $D(S)^{-1}(\epsilon)$ $<$, $=$, $>$ $U(S)^{-1}(\epsilon)$
(see Figure~\ref{fig:envelopes}).
Since $\epsilon \in \epsinterval{)}$ this can be calculated as follows:
let $S=\cup_{j=1}^k I_j$ for some $k\geq 1$, where each $I_j$ is a binary interval.
Then $U(S)^{-1}(\epsilon) = \min_{j=1}^k U(I_j)^{-1}(\epsilon)$ and $D(S)^{-1}(\epsilon) = \max_{j=1}^k D(I_j)^{-1}(\epsilon)$.
For a binary interval $I$, to determine $U(I)^{-1}(\epsilon)$, and similarly $D(I)^{-1}(\epsilon)$, go through the segments of its bounded envelope until the segment $r$ containing $\epsilon$ is found.
This search
alternates back and forth starting at the topmost and bottommost essential segments.
This is only performed during a feasibility test, which will result in either the segments above $r$, or those below $r$, becoming inessential.
Thus the time to find $r$ is at most a constant plus a term linear in the number of segments that become inessential.
Here too the linear term is charged to the inessential segments.

Any interval \inter{i}{j} can be decomposed into $\leq 2 \lfloor \lg (j\!-\!i\!+\!1) \rfloor + 1$
binary intervals where the sizes increase and then decrease, with perhaps two intervals of the same size in the middle.
E.g., $\inter{2}{13} =$ $\inter{2}{2} \cup \inter{3}{4} \cup \inter{5}{8} \cup 
\inter{9}{12} \cup \inter{13}{13}$.
These can be visited in $O(\log n)$ time by a tree traversal starting at the leaf $\inter{i}{i}$, moving upward to the least common ancestor of $i$ and $j$, and then downward to the leaf $\inter{j}{j}$.
Suppose, given $i$ and $\epsilon \in \epsinterval{)}$, we want to find the largest $j$ such that the error of making \inter{i}{j} a single step is $\leq \epsilon$.
We do this by a traversal to locate $j+1$.
By incrementally updating the $\min$ values to determine $U^{-1}_S(\epsilon)$, and $\max$ values used for $D^{-1}_S(\epsilon)$, when moving upward at node $p$ one can determine if $j+1$ is in $p$'s subtree (and hence the traversal should start going downward) by using $p$'s envelopes to decide if adding the entire subtree gives an error $> \epsilon$.
When moving downward, $j+1$ is in $p$'s left subtree iff adding the left subtree gives error $> \epsilon$, otherwise it is in $p$'s right subtree.
Not counting the queries of children's envelopes, the nodes visited are the same as those in going from $i$ to $j+1$ when $j+1$ is known.

Given $b$ and $\epsilon \in \epsinterval{)}$, a \textit{feasibility test} determines if there is a $b$-step function with regression error $\leq \epsilon$.
This can be accomplished by starting at 1 and determining the largest $j_1$ for which the error of making \inter{1}{j_1} a single step is $\leq \epsilon$, then starting at $j_1+1$ and determining the largest $j_2$ for which the error of making \inter{j_1+1}{j_2} a single step is $\leq \epsilon$, etc.
If the $b^\mathrm{th}$ step is finished before $n$ is reached then $\epsilon$ is infeasible, the test stops, and $\epslow=\epsilon$.
Otherwise, $\epsilon$ is feasible, the steps have been identified, and $\epshigh=\epsilon$.

To count the number of nodes visited, for each step the traversal visits nodes at a given height at most twice, once moving upward and once moving downward.
Thus at any height at most $2b$ nodes are visited.
The top $\lfloor \lg b \rfloor$ levels have a total of $\Theta(b)$ nodes.
There are $\lceil \lg n \rceil - \lfloor \lg b \rfloor = \Theta(\log n/b)$ lower levels, so in total 
$\Theta(b(1+\log n/b))$ nodes are visited.
Each visit takes $\Theta(1)$ time, so this is also the time required.

\subsection{Constructing the Tree}

We reduce the time to construct the interval tree of bounded envelopes by continually shrinking \epsinterval{]}.
At the end, \epsinterval{]} is so small that each bounded envelope is a single segment.
See Figure~\ref{fig:constructingtree}.

\begin{figure}
\setlength{\Ainindent}{0.4in}

\noindent
\textsf{\small
\hspace*{-0.08in}
\Ain{0} initialize envelopes of leaf nodes, $\epslow=0$, $\epshigh=\infty$, R $= \emptyset$\\
\Ain{0} for h=1 to $\lg n$~ \{h is height\}\\
\Ain{1}    for every binary interval I at height h, make I's envelopes by merging children's envelopes,\\
\Ain{2} and add any segment endpoint errors in \epsinterval{)} to R\\
\Ain{1} repeat 3 times~~\{reducing $|R|$ to $< n/2^{h}$ and total essential segments at height h $< 3n/2^h$\}\\
\Ain{2}    feasibility test using median of remaining essential segment endpoint errors in R\\
\Ain{0} do 2 more feasibility tests, reducing R to $\emptyset$, i.e., all envelopes are single essential segments
}

\caption{Constructing the Tree of Bounded Envelopes, Feasibility Tests Continually Shrink \epsinterval{]}} \label{fig:constructingtree}
\vspace*{-0.08in}
\hrulefill
\vspace*{-0.1in}
\end{figure}

First a feasibility test with $\epsilon = 0$ is performed using only the base level.
If it passes then the algorithm is done.
Otherwise, set $\epslow = 0$, $\epshigh = \infty$, and $R=\emptyset$.
Throughout, $R$ is an unordered multiset containing the errors of all segment endpoints (e.g., the error of the joint endpoint of \textsf{a} and \textsf{b} in Fig.~\ref{fig:envelopes}) remaining in \epsinterval{)}.
At height 0 each interval is a singleton, with single rays in its upward and downward envelopes for a total of $2n$ rays.
In general, at the end of height $h$, $|R| < n/2^h$ and the total number of essential segments in the envelopes at height $h$ is $< 3n/2^h$.
Moving upward, bounded envelopes from height $h$ are merged to form those at height $h+1$, creating
$ < 3n/2^{h}$ segments and $< 2n/2^h$ segment endpoints (the number of segment endpoints is the number of segments minus one per envelope).
Add the errors of the segment endpoints to those already in $R$, resulting in $|R| < 3n/2^h$.
Take the median error in $R$ and do a feasibility test.
Depending on the outcome, one of \epslow\ and \epshigh\ is adjusted and 1/2 the entries in $R$ can be eliminated.
Doing this 3 times results in $|R| < n/2^{h+1}$.
The number of essential segments is $\leq |R| + $ 1 per envelope (since $R$ included segment endpoints at level $h+1$), and hence is $< 3n/2^{h+1}$.

When the top is finished $|R| \leq 2$ and 2 feasibility tests are used to eliminate the remaining endpoint errors, i.e., at every node of the interval tree the upward and downward bounded envelopes have only one segment.
Complete the tree construction by removing all inessential segments, taking $\Theta(n)$ time.
Feasibility tests during tree construction have a slight change from standard traversals in that when height $h$ is being constructed, when the test's traversals reach height $h$ they go sideways, not upwards, from one node to the next since nodes at higher levels haven't yet been constructed.
This increases the total number of nodes visited per test by at most $n/2^h$.
Since only 3 tests are done per height (see Figure~\ref{fig:constructingtree}), this adds $\Theta(n)$ total time over all heights.
Thus the total time to construct the tree is $\Theta(n + \log n \cdot b(1 + \log n/b))$.

\subsection{Search for Minimal Feasible Error} \label{sec:SortedMatrix}

The $L_\infty$ error of a stepwise approximation is the maximum of the $L_\infty$ errors of its steps,
thus there is an interval $\inter{i}{j}$ such that the error of an optimal $b$-step approximation is the error of using the weighted $L_\infty$ mean as the step value on $\inter{i}{j}$.
Thus searching through such errors can determine the minimal feasible error.
``Parametric search'' was used in~\cite{FournierVigneronLinftyParametric,JQReducedIso_IF2012} but this is only of theoretical interest since parametric search is completely impractical, involving very complex data structures and quite large constants.

\textit{Search in a sorted matrix} provides a practical approach~\cite{FredericksonJohnsonSortedMatrix}.
Let $E$ be the $n \times n$ matrix where $E(i,j)$ is the error of using the $L_\infty$ mean on \inter{i}{j} if $i \leq j$, and is 0 if $i > j$.
$E$ is not actually created, but serves as a conceptual guide.
Few of its entries are ever determined.
Its rows are nondecreasing and the columns are nonincreasing, so
for any submatrix its largest entry is in the upper right and the smallest is in the lower left.

The algorithm has stages $0 \ldots \lg n -\!1$, where at the start of stage $s$ there is a collection of disjoint square submatrices of size $n/2^s$.
Stage 0 starts with all of $E$.
At each stage, divide each of the matrices into quadrants, and let $\epsilon_1$ be a median of the smallest value from each quadrant and $\epsilon_2$ a median of their largest values.
Values determined to be outside \epsinterval{)}, as in Figure~\ref{fig:envelopes}, are not calculated exactly and are set to $\epslow-1$ or $\epshigh+1$, as appropriate.
Feasibility tests for $\epsilon_1$ and $\epsilon_2$ are done, resulting in improvements to \epslow\ and/or \epshigh.
Quadrants with smallest value $\geq \epshigh$, or largest value $\leq 
\epslow$, are eliminated.
The remaining quadrants are the matrices that start the next stage.
Note that if $\epsopt < \epshigh$ then any quadrant with an entry of \epsopt\ is not eliminated,
hence at the end of each stage, either $\epsopt = \epshigh$ or one of the entries in the remaining submatrices is \epsopt.

After the last stage the remaining matrices are $1 \times 1$ and a standard binary search on these values is used to finish the determination of \epsopt.
The search uses $\Theta(\log n)$ feasibility tests, and the less obvious fact, proven in~\cite{FredericksonJohnsonSortedMatrix}, is that only $\Theta(n)$ entries of $E$ are evaluated.

\subsection{Evaluating $E$\, During The Search}

\begin{figure}

\begin{center}
\resizebox{4.5in}{!}{\includegraphics{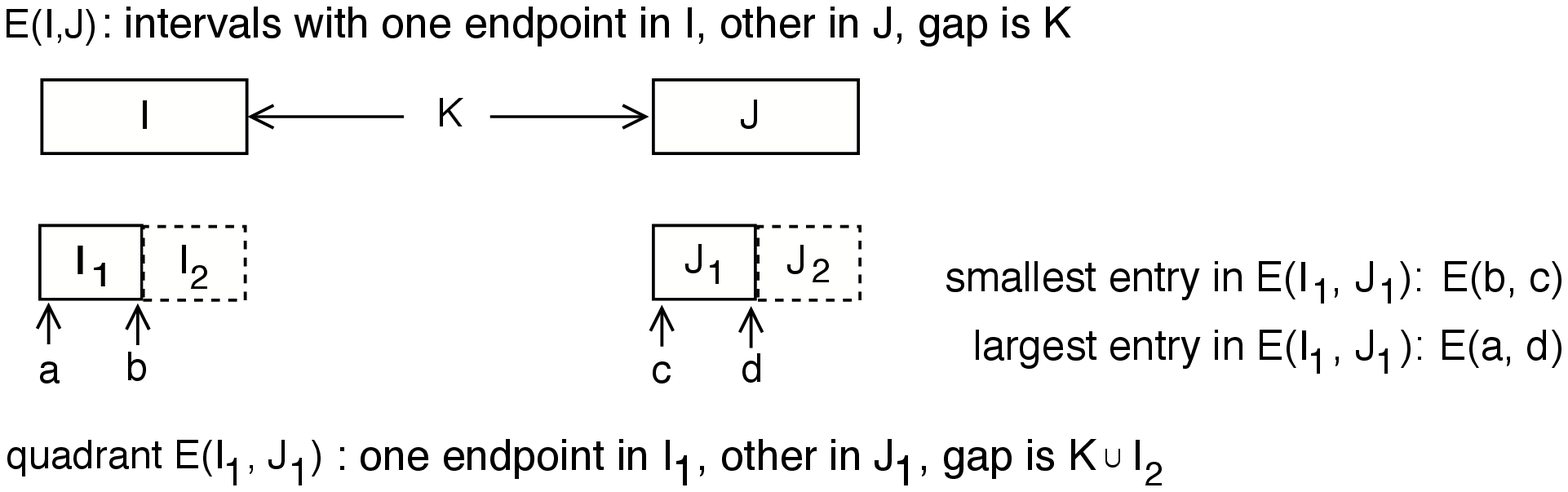}}
\end{center}
\vspace*{-0.2in}

\caption{Evaluating smallest, largest entries in E($I_1$, $J_1)$ using gap from $E(I,J)$}  \label{fig:Eexplain}
\vspace*{-0.08in}
\hrulefill
\vspace*{-0.15in}
\end{figure}

For intervals $I, J \subseteq \inter{1}{n}$ let $E(I,J)$ denote the submatrix $\{E(i,j): i\in I, j\in J\}$,
i.e., the submatrix corresponding to intervals starting at some $i \in I$ and ending at some $j \in J$.
At the start of stage $s$ of the search there is a collection of submatrices of the form $E(I,J)$ for binary intervals $I$, $J$ of size $n/2^s$.
Either $I = J$, or $I$ is to the left of $J$ and there is a (perhaps empty) gap between them with length an integral multiple of $n/2^s$ ($K$ in Figure~\ref{fig:Eexplain}).
During stage $s$, the quadrants of $E(I,J)$ are formed by cutting $I$ and $J$ in half into $I_1$, $I_2$ and $J_1$, $J_2$, respectively, creating quadrants $E(I_1,J_1)$, $E(I_1,J_2)$, $E(I_2,J_1)$, and $E(I_2,J_2)$.
The smallest and largest value in each quadrant needs to be determined, and as Figure~\ref{fig:Eexplain} shows, one can evaluate the smallest entry in $E(I_1,J_1)$ (i.e., $E(b,c)$), by using the envelopes from $K$ and the binary intervals \inter{b}{b}, $I_2$, and \inter{c}{c}.
The largest entry in the quadrant, $E(a,d)$, uses envelopes from $K$ and the binary intervals $I$ and $J_1$.
Similar results hold for all of the other quadrants of $E(I,J)$.
Exact values for entries outside \epsinterval{]} are irrelevant and $\epslow -1$ or $\epshigh +1$ is used, as appropriate.

The bounded envelopes for gap $K$ are associated with $E(I,J)$, and if, say, $E(I_1,J_1)$ is kept for stage $s+1$ then the envelopes for $I_2 \cup K$ are associated with it.
Just as for the tree construction, as search in a sorted matrix is proceeding the number of segments in the gap envelopes is reduced by interleaving feasibility tests based on the segment endpoint errors with tests for the basic search.
At the end of stage $s$ each gap envelope is copied at most 4 times, and each quadrant passed on to the next level adds $\leq$ 3 binary intervals which have envelopes that are only single segments.
As shown in~\cite{FredericksonJohnsonSortedMatrix} there are $\leq 2^{s+3}-1$ such quadrants, so by using 2 additional feasibility tests at each stage the total number of segments in the bounded envelopes at stage $s$ is $O(2^s)$.
Since the time to evaluate an entry of $E$ is linear in the number of segments involved, the total time to evaluate entries of $E$ over all stages of the algorithm is $\Theta(n)$, and the total time of the $\Theta(\log n)$ feasibility tests is $\Theta(\log n \cdot b(1 + \log n/b))$ plus the $\Theta(n)$ time to remove inessential segments.

This completes the proof of Theorem~\ref{thm:mainresult}.

\section{Final Comments}\label{sec:Final}

We have shown how to find an $L_\infty$ $b$-step approximation of weighted data, presorted by its independent coordinate,
in $\Theta(n + \log n \cdot b (1+\log n/b))$ time.
No previous algorithm~
\cite{ChenWangPiecewise2013,ChenWangnlogn2013,DiazBanezLinearDecide,FournierVigneronLinftyStep,FournierVigneronLinftyParametric,FulopPrillLinftyStep,GuhaShimLinftyHistogram,JQReducedIso_IF2012,KarrasetalHistogramDuality,LiuRandomizedLinftyReduced,MaysterLopezStep} 
was $o(n \log n)$ whenever $b = o(n)$, nor $\Theta(n)$ whenever $b = O(n/\log n \log\log n)$.
For sorted data the algorithm solves the 1-dimensional weighted $k$-center problem in the same time.

With a small change the algorithm also produces a ``reduced isotonic'' $b$-step function.
$f$ is an \textit{isotonic} function iff $f(1) \leq f(2) \leq \ldots \leq f(n)$, and is
an \textit{optimal $L_\infty$ $b$-step reduced isotonic regression of \data} iff it minimizes the $L_\infty$ error among all isotonic $b$-step functions.
Isotonic regression is an important form of nonparametric regression that allows researchers to replace parametric assumptions with weaker shape constraints~\cite{BarlowetalBook,RobertsonWrightDykstra}.
Some researchers were concerned that it can overfit the data and/or be too complicated~\cite{HaiminenetalReducedUnimodal,SalantiUlmReduced,SchellSingh} and resorted to reduced isotonic regression.
However, they used approximations because previous exact algorithms were too slow.
Merely changing the feasibility test to insure increasing steps finds $b$-step reduced isotonic regression in the same time bounds as $b$-step approximation.

\medskip
\noindent \textbf{Acknowledgement}: Research partially supported by NSF grant CDI-1027192


\begin{thebibliography}{99}


\bibitem{BarlowetalBook} Barlow, RE, Bartholomew, DJ, Bremner, JM, and Brunk, HD,
  \textit{Statistical Inference Under Order Restrictions: The Theory and Application of Isotonic Regression}, John Wiley, 1972.
\bibshrink
  
\bibitem{ChenWangPiecewise2013}  Chen, DZ and Wang, H,
  ``Approximating points by a piecewise linear function'',
  \textit{Algorithmica} 66 (2013), pp.\ 682--713.
\bibshrink

\bibitem{ChenWangnlogn2013} Chen, DZ and Wang, H,
  ``A note on searching line arrangements and applications'',
  \textit{Info.\ Proc.\ Let.}\ 113 (2013), pp.~518--521.
\bibshrink
  
\bibitem{DiazBanezLinearDecide}  D\'{i}az-B\'{a}\~{n}nez, JM and Mesa, JA,
  ``Fitting rectilinear polygonal curves to a set of points in the plane'',
  \textit{Eur.\ J. Operational Res.} 130 (2001), pp.~214--222.
\bibshrink
   
\bibitem{FournierVigneronLinftyStep} Fournier, H and Vigneron, A,
  ``Fitting a step function to a point set'',
  \textit{Algor.} 60 (2011), pp.~95--109.
\bibshrink
  
\bibitem{FournierVigneronLinftyParametric} Fournier, H and Vigneron, A,
  ``A deterministic algorithm for fitting a step function to a weighted point-set'',
  \textit{Info.\ Proc.\ Let.}\ 113 (2013), pp.\ 51--54.
\bibshrink
  
\bibitem{FredericksonJohnsonSortedMatrix}  Frederickson, G and Johnson, D,
  ``Generalized selection and ranking: Sorted matrices'',
  \textit{SIAM J.\ Comp.} 13 (1984), pp.~14--30.
\bibshrink
  
\bibitem{FulopPrillLinftyStep} F\"{u}l\"{o}p, J and Prill, M,
  ``On the minimax approximation in the class of the univariate piecewise constant functions'',
  \textit{Oper.\ Res.\ Let.} 12 (1992), pp.~307--312.
\bibshrink
  
\bibitem{GuhaShimLinftyHistogram}  Guha, S and Shim, K,
  ``A note on linear time algorithms for maximum error histograms'',
  \textit{IEEE Trans.\ Knowledge and Data Engin.}\ 19 (2007), pp.~993--997.
\bibshrink
    
\bibitem{HaiminenetalReducedUnimodal}   Haiminen, N, Gionis, A, and Laasonen, K,
  ``Algorithms for unimodal segmentation with applications to unimodality detection'',
  \textit{Knowl.\ Info.\ Sys.} 14 (2008), pp.~39--57.
\bibshrink

\bibitem{JQReducedIso_IF2012}   Hardwick, J and Stout, QF,
  ``Optimal reduced isotonic reduction'',
  \textit{Proc.\ Interface 2012}.
\bibshrink
  
\bibitem{KarrasetalHistogramDuality}  Karras, P, Sacharidis, D., and Mamoulis, N,
  ``Exploiting duality in summarization with deterministic guarantees'',
  \textit{Proc.\ Int'l.\ Conf.\ Knowledge Discovery and Data Mining} (KDD) (2007),
  pp.\ 380--389.
\bibshrink
  
\bibitem{LiuRandomizedLinftyReduced} Liu, J-Y,
  ``A randomized algorithm for weighted approximation of points by a step function'',
  \textit{COCOA} 1 (2010), pp.\ 300--308.
\bibshrink
  
\bibitem{MaysterLopezStep}  Mayster, Y and Lopez, MA,
  ``Approximating a set of points by a step function'',
  \textit{J. Vis.\ Commun.\ Image R.} 17 (2006), pp.\ 1178--1189.
\bibshrink
  
\bibitem{RobertsonWrightDykstra}  Robertson, T, Wright, FT, and Dykstra, RL,
  \textit{Order Restricted Statistical Inference}, Wiley, 1988.
\bibshrink
  
\bibitem{SalantiUlmReduced}  Salanti, G and Ulm, K,
  ``A nonparametric changepoint model for stratifying continuous variables under order
  restrictions and binary outcome'',
  \textit{Stat.\ Methods Med.\ Res.}\ 12 (2003), pp.\ 351--367.
\bibshrink

\bibitem{SchellSingh}  Schell, MJ and Singh, B,
  ``The reduced monotonic regression method'',
  \textit{J.\ Amer.\ Stat.\ Assoc.}\ 92 (1997), pp.\ 128--135.


\end{thebibliography}
\end{document}